\documentclass[aps,prl,showpacs,twocolumn,groupedaddress]{revtex4}

\usepackage{graphicx,color}
\usepackage{bm, amsmath, amssymb}

\begin{document}

\title{ Decoupled Spin Chains in Ba$_3$CuSb$_2$O$_9$}

\author{Lara Thompson$^{1}$ and Patrick A. Lee$^{1}$}

\affiliation{$^{1}$Department of Physics,
Massachusetts Institute of Technology,
Cambridge, MA 02139}


\begin{abstract}
Inspired by recent measurements on Ba$_3$CuSb$_2$O$_9$, we explore the interplay of spin and orbital degrees of freedom in a triangular lattice for possible spin liquid-like behaviour. Our effective Hamiltonian describes the magnetic Cu layer by considering superexchange paths along the intermediate O's. A mean field study of ordered states reveals an orbitally ordered ground state that isolates   the spin degrees of freedom   into spin chains. This scenario naturally explains the linear specific heat as $T\to 0$ and, by including the effects of additional defect  Cu's sitting off the triangular planes,  we find remarkable agreement with the measured magnetic susceptibility. 
\end{abstract}

\maketitle

The quantum spin liquid is an exciting and heavily sought ground state in which frustrated interactions lead to novel quantum order. In 1D, the antiferromagnetic (afm) spin chain has a spin liquid ground state and, indeed, it is well understood  and has been realized experimentally \cite{BonnerFisher,EgAff1992,SorEgAff1993}; however, the real excitement is in finding a gapless two dimensional quantum spin liquid. The prototypical frustrated system in 2d is the afm triangular lattice: with isotropic exchange, however, the ground state is not a spin liquid, but rather, it stabilizes into the 120$^{\circ}$ ordered state \cite{isotri}. There exist  numerous candidate quantum spin liquids \cite{balents}, including a variety of organic compounds and, more recently, the inorganic Ba$_3$CuSb$_2$O$_9$ \cite{balicas} and its sister compound Ba$_3$NiSb$_2$O$_9$ \cite{balicasNi}. The magnetic ions in these compounds, the spin 1/2 Cu and spin 1 Ni, respectively, form a triangular lattice in planes that are effectively isolated by two layers of the non-magnetic Sb. Recent measurements on the Cu system provide evidence that the low temperature state of Ba$_3$CuSb$_2$O$_9$ may be a quantum spin liquid. 
Motivated by these results, we seek to explain the spin liquid behaviour by examining the superexchange in the triangular lattice including the two-fold orbital degeneracy of the Cu.

The geometric frustration of the triangular lattice may be enhanced by the competition of the spin and orbital degrees of freedom \cite{mila2007, normand2008}; however, the coupling with the orbital degree of freedom often relieves the frustration via  orbital ordering \cite{orbord,LeeTiOCl}. Indeed, our analysis predicts a constant orbital ordering that entirely decouples the spin degrees of freedom on the triangular lattice into afm spin chains. 1D spin chains tend to a linear specific heat at low temperatures and, furthermore, the magnetic susceptibility does not show a conventional magnetic ordering (but, rather, has a distinctive maxima at $k_BT \approx J_{afm}$ \cite{BonnerFisher}). When we account for the impurity Cu substitutions into neighbouring layers of Sb, we naturally reproduce the temperature dependence of the magnetic susceptibility. We predict a defect substitution of Sb by 8.5\% Cu (more than the  5\% Cu substitutions reported in \cite{balicas}), coupling {\em ferro}magnetically with the nearest Cu in the triangular lattice, while the in-chain Cu's couple antiferromagnetically with an exchange $J_{afm} \approx 60$ K, to be compared  with the reported 32 K extracted for a triangular lattice \cite{balicas}.  

The Cu ions sit within an octahedron of O's, tilted relative to the neighbouring octahedra, so that the single d hole lies in the degenerate e$_g$ orbital doublet. 
Superexchange most likely occurs with an intermediate double occupancy of holes of an O$^{2-}$ to a neutral state. Virtual  hopping proceeds along pairs of O  in nearest proximity of the neighbouring tilted octahedra. The hybridization energy of Sb is much higher so that they aren't expected to participate.

\begin{figure}
\begin{center}
\includegraphics[width=3.2in]{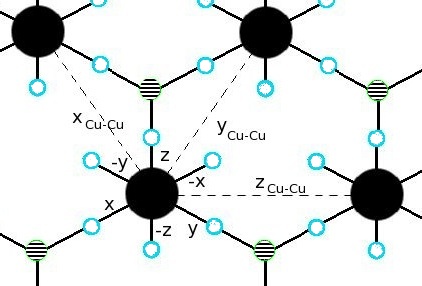}
\caption{\label{fig:bondnaming} A single Cu (solid) layer and the adjacent Sb(1) (striped) layer  -- the adjacent Sb(2) layer resides directly under the Cu triangular lattice. The O (hollow) reside midway between each Cu-Sb. The axes x, y and z form a right-angled coordinate system: the Sb(1)'s are drawn in the layer above the Cu layer so that the labelled local axes come diagonally out of the page to three such Sb(1)'s; the negative Cartesian directions are toward the O's sitting at right angles to the Sb(2) beneath each Cu. 
}
\end{center}
\end{figure}

Figure \ref{fig:bondnaming} depicts the triangular lattice of Cu's within a cage of O's connected to the inequivalent up versus down Sb layers. The e$_g$ orbitals are defined relative to the (arbitrarily chosen) $z$-axis and will be denoted d$_{3z^2-r^2}$ (or more simply just d$_{z^2}$), or by the spinor $\tau_z = +1$, and d$_{x^2-y^2}$, corresponding to $\tau_z=-1$. The spin is denoted by $\bf s$. For hopping along two intermediate O's directed along the $a$ and $b$  axes, the corresponding   Cu-Cu `bond'  is labelled  $c \neq a,b$ where $a,b,c\in \{x,y,z\}$.  The hopping into the upward plane is along an ordered pair of positive axes, say $(a,b)$, while the alternative hopping connecting the same Cu-Cu pair into the downward plane is along $(-b,-a)$.

Assuming the full octahedral symmetry in the Cu-O arrangement, 
  the intermediate hopping integrals share the following relations \cite{symarg}: 
\begin{align}\label{hoppingrelns}
t_{d_{z^2} p_z} = 2 t_{d_{z^2}p_{x/y}} = \pm {2\over \sqrt{3}} t_{d_{x^2-y^2}p_{x/y}}
\end{align}
A trigonal distortion,  stretching or compressing along the (111) direction, would not lift the e$_g$ orbital degeneracy; however,  these relations would become merely approximate.
Allowing for inequivalent energetics and hopping in the two Sb layers, we define two overall exchange parameters $J_1$ and $J_2$ for exchange via the Sb(1) and Sb(2) layers, respectively. Since the hopping is essentially longitudinal (the d and p orbitals are roughly aligned), we consider double occupancy of the same p orbital and the superexchange is antiferromagnetic.

The superexchange path between layers of Cu differs from the nearest neighbour intra-layer exchange path by additional hopping along a third O; the orientations of the hopping along two of the three O's is identical that of the hopping involved in the intra-layer superexchange. We can conclude that the inter-layer exchange is strictly smaller than the intra-layer exchange (by two orders in perturbation theory). As an aside, we point out that although the isostructural 6H-B Ni-based compound has no orbital degeneracy -- it has been lifted by the strong Hund's coupling that leads to a spin 1 degree of freedom -- the same conclusion holds for it:  the intra-layer exchange should be strictly larger than the inter-layer exchange, in disagreement with a recent proposal \cite{colorado}.

The resulting effective Hamiltonian is
\begin{align}\label{Heff}
H_{eff} = \sum_i  {\bf s}_i\cdot {\bf s}_{i+y}& \left( J_1Y[\tau_i, \tau_{i+y}]  +J_2 Y[\tau_{i+y}, \tau_{i}]    \right) \nonumber\\
+ {\bf s}_i\cdot {\bf s}_{i+x} &\left( J_1X[\tau_i, \tau_{i+x}]  +J_2X[\tau_{i+x}, \tau_{i}]    \right) \nonumber \\
+ {\bf s}_i\cdot {\bf s}_{i+z} &\left( (J_1+J_2) Z^S[\tau_{i},\tau_{i+z}]  \right. \\
& \left. -\sqrt{3}(J_1-J_2  )Z^A[\tau_{i},\tau_{i+z}] \right) \nonumber
\end{align}
where the orbital dependence of the various bonds is
\begin{align}
Y/X[ \tau_{i+y/x},\tau_i] &= { \tau_{i+y/x}^z+1\over 2}  \left[ 4(2 - \tau_{i}^z) \pm 4 \sqrt{3}\tau_{i}^x\right] \\
Z^S[\tau_{i+z},\tau_i] &= \tau_{i}^z\tau_{i+z}^z - 2(\tau_i^z+\tau_{i+z}^z) + 4 - 3 \tau_{i}^x\tau_{i+z}^x \\
Z^{A}[\tau_{i+z},\tau_i] &= (2-\tau_i^z)\tau_{i+z}^x - \tau_i^x (2-\tau_{i+z}^z) 
\end{align}
Although we artificially break the full octahedral symmetry in our choice of e$_g$ basis, this effective Hamiltonian retains the full symmetry that can be verified by a rotation of the orbital basis, such as $(x,y,z) \to (z',x',y')$ which corresponds to a 60$^\circ$ rotation of the orbital spinor $\tau$. 

We analyze the effective Hamiltonian by assuming a mean field decoupling of the spin and orbital degrees of freedom. The spin and orbital directions are specified by a pair of polar winding angles $\alpha_j$ and $\beta_j$, respectively:
\begin{align}
{\bf s}_i &= s ( \cos(\theta_s + {\bm \alpha} \cdot {\bm a}_i), e^{i\phi_s + {\bm \alpha}' \cdot {\bm a}_i}\sin(\theta_s + {\bm \alpha} \cdot {\bm a}_i)) \label{spin}\\
\tau_i &= (\cos (\theta_\tau +{\bm \beta} \cdot {\bm a}_i), e^{i\phi_\tau + {\bm \beta}' \cdot {\bm a}_i} \sin(\theta_\tau +{\bm \beta} \cdot {\bm a}_i) )\label{tau}
\end{align}
 The mean field solutions do not depend on the azimuthal angles, $\phi$, $\alpha'$, or $\beta'$: we consider rotations about the $y$ axis without loss of generality. Note that the mean field energy  never depends on the angle $\theta_s$ but depends on $\theta_\tau$ in the special cases of `ferromagnetic (fm)', $\beta = 0$, or `afm', $\beta = \pi$, (or alternating) orbital ordering.

The mean field ground state corresponds to a `fm' orbital ordering in the $d_{x^2-y^2}$ orbital that decouples the spin degree of freedom into afm spin chains along the $z$ axis.  This can be seen immediately from the Cu-O arrangement depicted in Figure \ref{fig:bondnaming}: exchange coupling along the Cu-Cu $x$ and $y$ bonds inevitably involves hopping along the $z$ axis along which the planar  $d_{x^2-y^2}$ orbital has zero overlap. 
 There are also degenerate solutions corresponding to chains along the $x$ and $y$ axes that are most easily constructed by a coordinate rotation in orbital space. The afm chain exchange is completely independent of the up/down asymmetry: indeed, $J_{afm} = 9(J_1+J_2)$.  No other mean field orbital ordering described by (\ref{spin}-\ref{tau}) locally minimizes the energy. 

A trigonal distortion modifies our exact relations amongst the hopping integrals (\ref{hoppingrelns}) and hence also modifies the precise numerical prefactors in the effective Hamiltonian (\ref{Heff}). However, we expect our conclusions will remain unchanged. This is because we can rotate to a new basis for the e$_g$ orbitals where the new `planar' orbital is chosen such that the overlap with the p orbital along the $\tilde z$-axis vanishes in the distorted octahedron.  Then we are assured that an orbital ordering into this orbital will still break the triangular lattice into independent afm spin chains. 

Another possible distortion of the lattice  common in afm spin chains is a Peierls distortion which results in alternating short and long Cu-Cu `bonds' and favours singlet pairing of spins along the short `bond'. This sort of distortion would mix the e$_g$ orbital states and our simple symmetry arguments break down. Without a quantified analysis of the multiple exchange paths incorporating such deformations, we cannot comment on possibility of such a transition.

In the samples of Ba$_3$CuSb$_2$O$_9$,  Zhou et. al. report that the Cu impurities are substituted specifically onto the Sb(2) sites  \cite{balicas}.  Such an impurity Cu is connected via three alternative O's (in Figure \ref{fig:bondnaming}, these O's correspond to each of the negative axes) to the nearest in-plane Cu: the angle formed between either Cu and the intermediate O is a right angle (or nearly so) so that the exchange between the two is ferromagnetic \cite{KugelKhomskii}. At low energies, these fm coupled Cu's will form an effective spin 1 defect. The resulting effect on the remaining afm chains now depends on the effective exchange of this spin 1 with the next-nearest Cu's on either side in the chain. Note that since we must consider  additional exchange paths directly between the defect Cu and these next-nearest neighbour Cu's, we cannot conclude a priori if the resulting exchange will be antiferro- or ferromagnetic.  
With an afm exchange, the two next-nearest neighbour Cu's generically screen the spin 1 defect leaving no orphan spin and a cut spin chain less three spin sites altogether \cite{EgAff1992}; possible fine-tuning of the afm exchange leads only to partial screening of the spin 1 to an orphan spin 1/2 whilst healing the spin chain so that it remains uncut \cite{SorEgAff1993}. A fm exchange with the two chain ends renormalizes to an orphan spin 1, cutting the chain in two with one less spin. 

In order to discern our most likely defect scenario, consider the source of $1/T$ divergences in the magnetic susceptibility. An orphan spin $S$ contribution follows the simple Curie law: $\chi_{S} = (g\mu_B)^2S(S+1)/3k_B T$. The same density of orphan spins for spin 1 versus spin 1/2 therefore contributes a factor 8/3 larger $1/T$ divergence in the magnetic susceptibility. Finite odd length afm chains also diverge as $T\to0$; however, the region of $1/T$ behaviour is limited to $k_BT\lesssim 0.15J_{afm}$ \cite{BonnerFisher}. This difference in temperature scale allows us to separate the orphan spins and the broken chain contributions to the divergence. As a first approximation, we can  extract the defect concentration ignoring the chain contribution: for orphan spin 1, we require a concentration of 8.5\% defects, compared with $\sim$ 20\% spin 1/2 orphan spins. Although even this smaller estimate exceeds the reported 5\% Cu substitutions, we will hereafter assume the spin 1 orphan scenario, which, recall, also entails breaking the chains at the site of each defect.

The broken afm chain contribution is calculated assuming a random distribution of defects. The corresponding  probability that a Cu on the triangular lattice should find itself in a chain of length $n$ is 
\begin{align}\label{Pxn}
P_x(n) = nx^2(1-x)^n
\end{align}
where $x$ is the density of defects \cite{LeeTiOCl}. We fit the experimental magnetic susceptibility by combining the contributions of fm coupled pairs of spin 1/2 defects and broken afm chains varying the density of defects and both the chain afm exchange and fm defect exchange. The finite chain susceptibility and specific heat was calculated by exact diagonalization and interpolated in $1/n$ for large $n$ using the infinite chain Bonner-Fisher curve \cite{BonnerFisher}.

\begin{figure}[tbp]
\begin{center}
\includegraphics[width=0.5\textwidth]{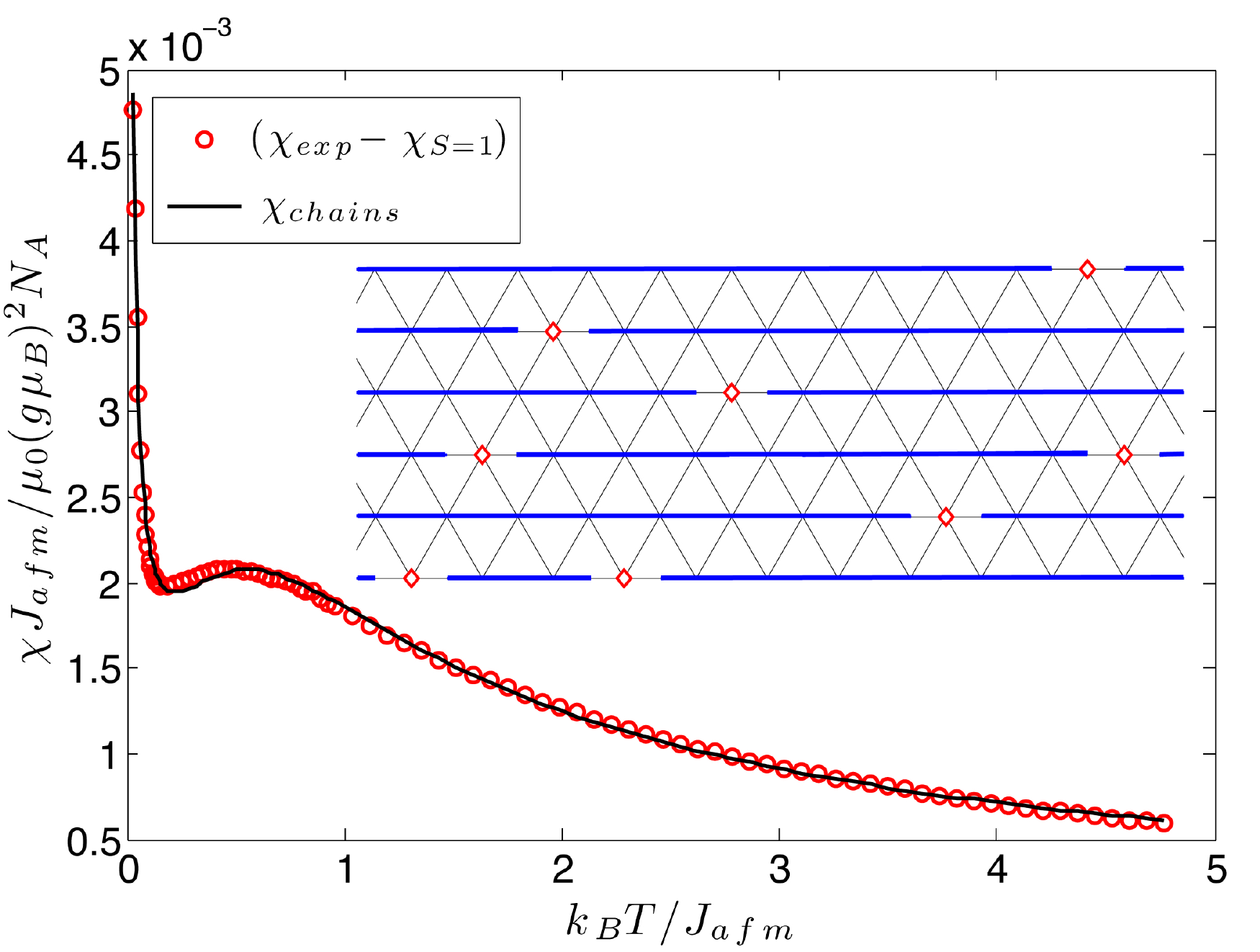}
\caption{\label{fig:chichain-like} A comparison of the experimental n$\chi$ (after orphan $S=1$ spin substraction) with the modelled broken afm spin chain response $\chi_{chains}$. Note that the $T\to0$ divergence is now due entirely to the odd length finite chains. The fit parameters involved a Cu defect substitution on Sb(2) sites of 8.5\% with fm exchange with the nearest Cu in-plane of  $J_{fm} = 34$ K, and a chain afm exchange of $J_{afm} =   63$ K. Inset: a graphical depiction of the triangular lattice decoupled into broken spin chains with spin 1 orphans at each substituted Cu defect.
}
\end{center}
\end{figure}

The fit of our model with the experimental data is best shown by subtracting the modelled orphan spin contribution  and comparing $\chi_{exp} - \chi_{S=1}$ with the broken chain susceptibility (see Figure 
 \ref{fig:chichain-like})\footnote{In the published data, only 5\% spin 1/2 orphan spins were subtracted. The resulting curve retains a large Curie component that obscured the hump in Figure \ref{fig:chichain-like}.}. The fit parameters are a  fm  exchange of $J_{fm}=34$ K coupling the 8.5\% substituted Cu's with the nearest in-plane Cu,   and an afm exchange of $J_{afm} = 63$ K within the broken chains. After subtracting the orphan spin contribution,  the $T\to0$   divergence is now due solely to the odd length broken chains.  The height and position of the local maxima are essentially determined by the afm exchange within the chains. The fm exchange affects the magnetic susceptibility curve far less drastically: in practice, all three parameters, $x, J_{afm}$, and $J_{fm}$ are fit simultaneously in a least squares minimization scheme.

\begin{figure}[tbp]
\begin{center}
\includegraphics[width=0.5\textwidth]{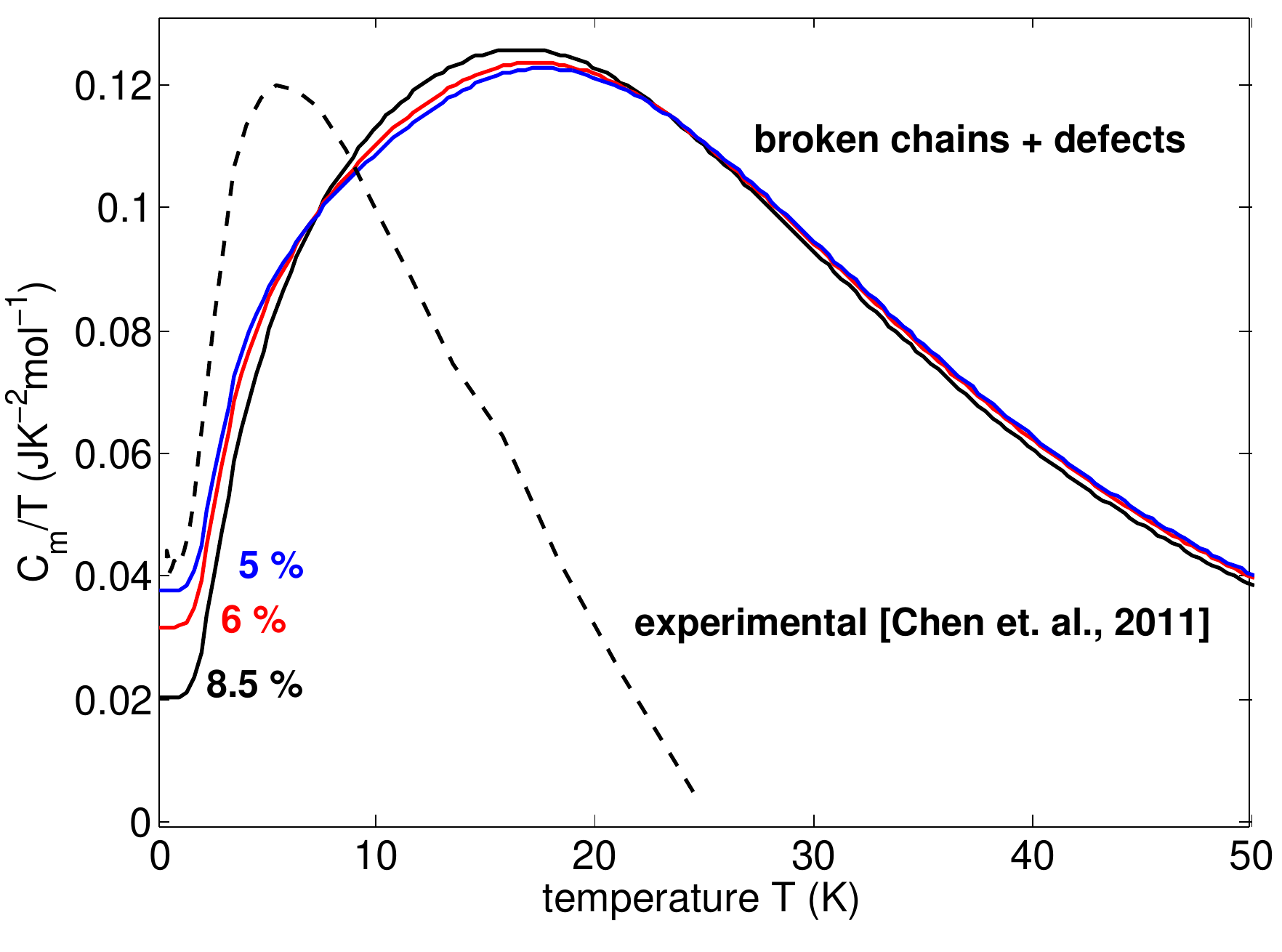}
\caption{\label{fig:Cchain-like} Using the best fit parameters to the susceptibility [$x=\{5, 6, \bm{ 8.5} \%\}$, $J_{fm} = 34$ K, $J_{afm} =   63$ K], the modelled specific heat of the broken chains and spin 1 defects. Note that the experimental intercept of $C_m/T$ corresponds to a defect concentration of roughly $x=4.5 \%$ in our model.}
\end{center}
\end{figure}

With these same fit parameters, we can compare our model's predictions against the experimental magnetic heat capacity. 
We reproduce the same linear behaviour as $T\to0$; the modelled peak in $C_m/T$ has the correct amplitude, however,  it is broader and persists to higher temperatures.  At low temperatures, the predicted intercept of $C_m/T \to \gamma$ decreases rapidly with increased defect density: the experimental intercept corresponds to a defect concentration of $x=4.5\%$ in our model. Since the linear specific heat arises from the long chain contributions, alternatively, a modified  chain length distribution (\ref{Pxn}) that favours longer chains will match the experimental intercept in $C_m/T$ with a higher defect concentration.  This would correspond to a repulsion between defect sites. The position of the peak and overall scaling of $C_m/T$ are determined primarily by the afm exchange in the Cu chains  and would be largely unaffected by any correlations between defect positions.
The measured heat capacity requires first a subtraction of the phonon contribution: its form was taken from the heat capacity of the non-magnetic isostructural compound Ba$_3$ZnSb$_2$O$_9$  \cite{balicas}. Above $T \sim 10$ K, the phonon contribution overwhelms the magnetic response, and the result after subtraction may be unreliable; indeed, the experimental curve extrapolates rapidly to negative values in that regime (the dashed curve in Figure \ref{fig:Cchain-like}).

In summary, we propose that the magnetic degrees of freedom in Ba$_3$CuSb$_2$O$_9$ decouple into antiferromagnetic spin chains due to a constant planar e$_g$ orbital ordering. The defect Cu substitutions on Sb(2) sites couples to an in-plane Cu to form effective spin 1 orphans that simultaneously break the spin chains at each defect site. Our modelled magnetic susceptibility reproduces the experimental curve remarkably well; however, the magnetic specific heat initial slope may require further consideration of correlations among the off-plane Cu defects. The predicted orbital ordering should to lead to an orthorhombic distortion of the Cu-O octahedral structure and should be discernible in further structural analyses of the compound. 

We thank H. D. Zhou, E. S. Choi, Luis Balicas and their collaborators for sharing their experimental data and for useful discussions. The work of L.T. is supported by NSERC. P.A.L. acknowledges  support of the NSF under grant no. DMR1104498.


\begin{thebibliography}{99}

  \bibitem{balicas} H. D. Zhou, E. S. Choi, G. Li, L. Balicas, C. R. Wiebe, Y. Qiu, J. R. D. Copley, and J. S. Gardner, ``Spin liquid state in the $S=1/2$ triangular lattice Ba$_3$CuSb$_2$O$_9$,'' Phys. Rev. Lett. {\bf 106}, 147204 (2011).

\bibitem{balicasNi} J. G. Cheng, G. Li, L. Balicas, J. S. Zhou, J. B. Goodenough, C. Xu, and H. D. Zhou, ``High-pressure sequence of Ba$_3$NiSb$_2$O$_9$ structural phases: new S=1 quantum spin liquids based on Ni$^{2+}$,'' Phys. Rev. Lett. {\bf 107}, 197204 (2011).

  \bibitem{isotri} B. Bernu, P. Lecheminant, C. Lhuillier and L. Pierre, ``Exact spectra, spin susceptibilities, and order parameter of the quantum Heisenberg antiferromagnet on the triangular lattice,'' Phys. Rev. B {\bf 50}, 10048 (1994).
  
  \bibitem{balents} L. Balents, ``Spin liquids in frustrated magnets,'' Nature {\bf 464}, 199 (2010).
  
  \bibitem{orbord} See  D. I. Khomskii, ``Role of orbitals in the physics of correlated electron systems,'' Physica Scripta. {\bf 72}, CC8 (2005), and references therein.
    
  \bibitem{mila2007} F. Mila, F. Vernay, A. Ralko, F. Becca, P Fazekas and K. Penc, ``The emergence of resonating valence bond physics in spin-orbital models,'' J. Phys.: Condens. Matter {\bf 19}, 145201 (2007).
  
  \bibitem{normand2008} B. Normand, A. M. Oles, ``Frustration and entanglement in the t2g spin-orbital model on a triangular lattice: Valence-bond and generalized liquid states,'' Phys. Rev.  B {\bf 78}, 094427 (2008).
  
  \bibitem{LeeTiOCl} A. Seidel, C. A. Marianetti, F. C. Chou, G. Ceder, and P. A. Lee, ``S=${1\over 2}$ chains and spin-Peierls transition in TiOCl,'' Phys. Rev. B {\bf 67}, 020405(R) (2003).
  
  \bibitem{BonnerFisher} J. Bonner and M. Fisher, ``Linear magnetic chains with anisotropic coupling,'' Phys. Rev. {\bf 135}, A640 (1964).
  
  \bibitem{sawatzky} T. Sudayama, Y. Wakisaka, T. Mizokawa, H. Wadati, G. A. Sawatzky, D. G. Hawthorn, T. Z. Regier, K. Oka, M. Azuma, and Y. Shimakawa, ``Co-O-O-Co superexchange pathways enhanced by small charge-transfer energy in multiferroic BiCoO$_3$,'' Phys. Rev. B {\bf 83}, 235105 (2011).
  
  \bibitem{KugelKhomskii} K. I. Kugel and D. I. Khomskii, ``Crystal structure and magnetic properties of substances with orbital degeneracy,'' Sov. Phys. JETP {\bf 37} (4), 725(1973); ``The Jahn-Teller effect and magnetism: transition metal compounds,'' Sov. Phys. Usp. {\bf 25} (4), 231 (1982).
  
  \bibitem{EgAff1992} S. Eggert and I. Affleck, ``Magnetic impurities in half-integer-spin Heisenberg antiferromagnetic chains,'' Phys. Rev. {\bf B 46}, 10866 (1992).
  
  \bibitem{SorEgAff1993} E. S. Sorensen, S. Eggert and I. Affleck, ``Integrable versus non-integrable spin chain impurity models,'' J. Phys. A: Math. Gen. {\bf 26}, 6757 (1993).
  
  \bibitem{symarg} The hopping integral relations can be found by considering a basis rotation $x,y,z \to z',x',y'$ and noting the azimuthal symmetry of the $d_{3z^2-r^2}$ orbital.
  
  \bibitem{colorado} G. Chen, M. Hermele, and L. Radzihovsky, ``Frustrated quantum critical theory of putative spin-liquid phenomenology in 6H-B-Ba$_3$NiSb$_2$O$_9$,'' cond-mat/1201.2181 (2012).
  
\end{thebibliography}
\end{document}